\documentstyle[12pt]{article}
\begin{document}
\vspace*{4.5cm}
\begin{center}
\begin{large}\begin{bf}
 DECONFINEMENT OF CONSTITUENT
QUARKS AND THE HAGEDORN TEMPERATURE\end{bf}\end{large}
\end{center}
{}~\\
{}~\\
{}~\\
\hspace*{2.54cm}O.D.Chernavskaya and E.L.Feinberg
\\

\hspace*{1.54cm}P.N.Lebedev Physical Institute of the Russian Academy of

\hspace*{1.54cm}Sciences, Russia, 117924 Moscow, Leninsky prospect 53

\hspace*{1.54cm}E-mail: feinberg@td.fian.free.net

\hspace*{1.54cm}Fax:(007)(095) 938-22-51; 135-85-33.
\vspace*{1.5cm}
\\
{\bf INTRODUCTION}
{}~\\
\par
The overwhelming majority of theoretical papers on
phase transition of hadronic matter  $H$ to quark gluon plasma
($~H\leftrightarrow~QGP$) ignores constituent quarks $Q$
which we shall call below briefly {\it valons} (the term proposed
by R.Hwa). The present paper is fundamentally based on the
conception of those valons as real entities having the
same quantum numbers as current quarks and the mass which
was more than once calculated theoretically
as some 300 MeV and used for
explanation of experiments (e.g., see$^{\rm 1}$).
 The possibility of existence
of a special state of strong interacting matter, that of
{\it deconfined valon gas with still broken chiral
symmetry}, is discussed here.
\par
Such a possibility, i.e. appearance of the third, $Q$ phase
intermediate between $H$ and $QGP$ had been discussed already
in papers$^{\rm 2-5}$ (we apologize for possibly
missing some other papers). The idea may be traced back to E.Shuryak
who pointed out that deconfinement and restoration
of chiral symmetry might not coincide$^{\rm 6}$. However in the papers
mentioned above the result turned out pessimistic. E.g., on
the $\mu -T$ diagram ($\mu$ - chemical potential, $T$ -
temperature) possible $Q$ phase (allowing also for admixture
of pions necessary first of all as Goldstone particles) occupies
but a very small area at large $T$ and rather small $\mu$
(see e.g. Fig.1 taken from$^{\rm 5}$).The "constituent
quark phase is strongly suppressed"$^{\rm 2}$.
\par
Meanwhile the intuitive physical pattern$^{\rm 7,8}$ based on
the bag model ideology, although being extremely rough, points to
the quite a different possibility schematically shown in
Fig.2. In fact, a nucleus consists of nucleons $N$ which in
turn are bags containing valons, which are nothing else than
a massive clouds of strong interacting current quarks $q$,
antiquarks $\bar{q}$ and gluons $g$ with quantum numbers of a single
quark. If we compress the nucleus (Fig.$2a$) until nucleons
become close packed and fill up the entire space (Fig.$2b)$
they merge into a single bag with valons deconfined (still
with chiral symmetry broken) and freely moving within the
common bag (Fig.$2c)$. Further compression  leads to a new
situation, that of close packed and merging valons (Fig.$2d)$. Now
$q, \bar{q}$ and $g$ which earlier were confined within valons can
freely propagate within a common bag (Fig.$2e)$. Here they are
deconfined and massless, chiral symmetry is authomatically
restored and this very state is $QGP$.
\par
Thus we expect two phase transitions to take place: the one,
$H\leftrightarrow Q$, and (at higher density and/or
temperature) another,  $Q\leftrightarrow QGP$.
\par
In this paper we are going to show, in distinction to$^{{\rm 2-5}}$,
that with reasonable values of bag
parameters $B_{Q}$
for the valon phase and $B_{q}$ for $QGP$ there actually appears
possibility for a pattern shown in Fig.2, i.e. the
possibility of the overall $H \leftrightarrow QGP$ transition to proceed
only via the intermediate $Q$ phase.
\par
The first $(H\leftrightarrow Q)$ transition establishes limitations on
possibility of existence of normal hadrons. This is the reason
for interpreting it as the Hagedorn limit and to consider
the temperature of transition, i.e. of valon deconfinement, $T_{d}$,
as the Hagedorn temperature $T_{H}$. The second transition restores
chiral symmetry and is characterized by the temperature $T_{ch}$. Thus
the Hagedorn temperature appears to be not merely a very rough
approximation to chiral symmetry restoration/ breaking
temperature $T_{ch}$ as is frequently believed. It has its own
physical meaning of temperature of constituent quark
$(\equiv $valon) deconfinement: $T_{H}\equiv  T_{d}$.\\
{}~\\~\\
{\bf THERMODYNAMICAL FORMALISM}
{}~\\
\par
We are treating phase transitions following traditional
manner  $^{2-5}$, fundamentals of which were given
by Hagedorn${\rm ^9}$. As the first step we consider  the
partition function $Z^{0}(T,\mu ,V)$ depending on $T$, $\mu$,
and system  volume $V$ for each phase in the first
approximation (which means: for pointlike particles):
\begin{eqnarray}
&&{\rm ln} Z^0_j (T,\mu,V) = \frac{V}{T} \sum_i \Biggl\{
\frac{g^B_i}{6\pi^2} \int \frac{k^4 dk}{\sqrt{k^2+m^2_i}}
\frac{1}{exp\left(\frac{\sqrt{k^2+m^2_i}}{T}\right) - 1}+  \nonumber \\[-2mm]
&&\\[-2mm]
&&\frac{g^F_i}{6\pi^2} \int \frac{k^4 dk}{\sqrt{k^2+m^2_i}} \left[\
\frac{1}{\exp\left( \frac{\sqrt{k^2+m^2_i}-\mu_i}{T}\right)+1} +
\frac{1}{\exp\left( \frac{\sqrt{k^2+m^2_i}+\mu_i}{T}\right) +1} \right]\
\Biggr\} \,\,+ {\rm ln} Z^0_{vac} \nonumber
\label{1}
\end{eqnarray}

\noindent where $g^{B,F}_{i}$and $m_{i}$ are degeneracy factors
and masses for Bose
and Fermi $i$-th type particles respectively for each $j$-th phase
$(j$ designates hadronic $H$, valonic $Q$, and current $(QGP),~q$
phases respectively). ln$Z^{0}_{j}(T,\mu ,V)$ is a result of summation
over particles which play considerable role (under $T$ and $\mu $
in question) in the given phase. In particular, for hadronic
phase $H$ we have taken into account nucleons (protons and
neutrons, $M=940$ MeV), pions $(m_{\pi }=140$ MeV) and
$\Lambda $ hyperons  $(m_{\Lambda } \simeq 1150$ MeV).
The constituent phase $Q$ contains light valons $\equiv $
constituent quarks $(m_{u}{\simeq}m_{d}{\simeq} 320$  MeV),
some  admixture  of  strange valon quarks $(m_{s}{\simeq} 512$ MeV),
and again pions. The term ln$Z^0_{vac}$ stands here for $Q$ and $QGP$
phases and reflects effective interaction of quarks and gluons
with the QCD-vacuum. This interaction in $Q$ and $QGP$ phases is
assumed as
\begin{equation}
{\rm ln}Z^0_{vac,Q}(T,\mu ,V)= - \frac{V}{T} B_Q~, 
\label{2}
\end{equation}
\begin{equation}
{\rm ln}Z^0_{vac,q}(T,\mu ,V)= - \frac{V}{T} B_q 
\label{3}
\end{equation}
\noindent where $B_{Q}, B_{q}$ are some constant bag parameters
differing from each other (to be discussed below).
\par
Then by usual differentiations of $Z^{0}_{j} (T,\mu ,V)$ we obtain
for each phase energy density $\epsilon ^{(0)}_{j}$,
pressure $p^{(0)}_{j}$, and particle density $n^{(0)}_{j}$as
functions of internal parameters $T,\mu $,$V$.
\par
Having this as a basis and following the procedure
proposed$^{\rm 2-5}$ (~more exactly in$^{\rm 2}$~) we introduce mutual
interaction of particles by ascribing them hard core radii in
the Van der Waals manner. Namely, we substitute $n^{(0)}$ with
real particle density $n^{(r)}$ and then obtain "real" energy
density $\epsilon ^{(r)}$ and pressure $p^{(r)}$ for each of three phases
$(j=H,Q$,$QGP$):

\begin{equation}
n^r_j (T,\mu) = \frac{n^0(T,\mu)}{1+n^0(T,\mu)\ast v_j}~,
\label{4}
\end{equation}
\begin{equation}
\epsilon^r_j(T,\mu) = \frac{\epsilon^0(T,\mu)}{1+n^0(T,\mu)\ast v_j}~,
\label{5}
\end{equation}
\begin{equation}
p^r_j(T,\mu) = \frac{p^0(T,\mu)}{1+n^0(T,\mu)\ast v_j}~,
\label{6}
\end{equation}

\noindent where each of the parameters $v_{j}$ is hard core
volume characterizing $j$-th  phase particles:
\begin{equation}
v_N = \frac{4}{3}\pi r^3_N~,
\label{7}
\end{equation}
\begin{equation}
v_Q = \frac{4}{3}\pi r^3_Q~,                            
\label{8}
\end{equation}
\noindent current quarks and gluons are of course pointlike $( v_{q}=0)$.
\par
At given $T$ and $V$ of the system the most preferable
(i.e. stable) phase is the one having the largest pressure
$p_{j}$ and smallest $\mu _{j}$. At equilibrium of each two phases theire
pressures and effective chemical potentials are to be equal. Three types
of phase equilibrium curves are to be analyzed: the one corresponding to
the  {\bf valon deconfinement} transition ($H\leftrightarrow Q$)
\begin{equation}
\mu _{H}=3\mu _{Q}; \qquad p_{H}(T_{d},\mu_{H})=p_{Q}(T_{d},\mu_{H}/3)~,
\label{9}
\end{equation}
\noindent {\bf direct} transition ($H\leftrightarrow QGP$)
\begin{equation}
\mu _{H}=3\mu _{q}; \qquad p_{H}(T_{c},\mu_{H})=p_{q}(T_{c},\mu_{H}/3)~,
\label{10}
\end{equation}
\noindent and {\bf chiral} transition ($Q\leftrightarrow QGP$)
\begin{equation}
\mu _{Q}=\mu _{q}; \qquad p_{Q}(T_{ch},\mu_{Q})=p_{q}(T_{ch},\mu_{H}/3)~.
\label{11}
\end{equation}
\noindent The point of {\bf coexistence of all three phases} (i.e.
the {\it triple point}) is defined by the following conditions:
\begin{equation}
p_{H} (T^{\#},\mu^{\#}) = p_{Q}(T^{\#},\mu^{\#}) = p_{q}(T^{\#},\mu^{\#})
\label{12}
\end{equation}
\noindent Before turning to study of equilibrium conditions and phase
transitions let us discuss the chosen set of parameters.
{}~\\~\\~\\
{\bf BASIC PARAMETERS}
{}~\\
\par
There are two pairs of decisively important parameters:
firstly, particle sizes, e.g. average root square radii of
nucleons $N, r_{N}$, understood as $(\bar {r^2}_{N})^{1/2}$, and
valons, $r_{Q}$, which designates $(\bar{r^2}_{Q})^{1/2}$.
 Secondly bag pressures, $B_{Q}$ for the $Q$ phase and $B_{q}$ for $QGP$.
\par
Let us discuss sizes. They are used$^{{\rm 2-4}}$ to describe
interaction of particles in a simplified form by considering
them as black balls forming a Van der Waals gas. This means
that we approximate  their interaction potential $U(r)$ as $U =
\infty $ for $r<r_{c}$ and $U=0$ for $r>r_{c}, r_{c}$
being the core radius (in other calculations we have used also
some continuous potential $U(r)$). In${\rm ^5}$ the nucleon
interaction was instead described by two versions of the continuous $U(r)$
satisfactorily describing nuclear properties. It seems
sufficient in this paper to consider $N$ and $Q$ simply as
having cores$^{\rm 2,3}$. For nucleon radius and volume
$r_N$, $v_N$ we usually assume, as most resonable, the values:
\begin{equation}
r_N \simeq 0.8 {\rm fm}; \qquad  v_N \simeq 2.13 {\rm fm}^{\rm 3}
\label{13}
\end{equation}
\noindent (however sometimes we shall consider also $r_N=0.7$ fm). Since
the nucleus has the radius $R_A = r_{0}*A^{1/3}~(A$ - atomic weight)
with $r_{0} \simeq  1.10-1.15 $fm, the fraction of the nucleus volume
occupied by nucleons is
\begin{equation}
(r_{N}/r_A)^3 \sim 0.26 \div 0.38~.
\label{14}
\end{equation}
\noindent Thus it seems sufficient to compress a nucleus only 3-4
times to come to close packing of nucleons.
\par
The size of valons has been estimated$^{\rm 10}$ from
analysis of properties of various hadrons within the
Additive Quark Model (AQM); when expressed by the square root
radius $r_{Q} ($for $r_{N}=0.8$ fm) this estimate sounds as:
\begin{equation}
0.20 \leq r_{Q} \leq 0.36 {\rm fm}. 
\label{15}
\end{equation}
\noindent We shall concentrate on some middle resonable value,
$r_{Q}= 0.3 $fm, although other values within limitation
(15) will be also used (let us remark that estimating $r_{Q}$
within black ball valon model and AQM  from $QQ$ collision
cross section $\sigma_{QQ} = \pi (2r_{Q})^{2}\simeq (1/9)\sigma _{NN}$
where $\sigma _{NN}$ is the $NN$
collision cross section taken, at not too high energies, as
$40$ mb, we get $r_{Q}= 0.2$ fm. For tevatron energies it is much larger).\\
\par
Now, about $B_{Q}$ and $B_{q}$. The choice of their values is much more
disputable, being at the same time very essential. Previous
authors$^{\rm 2-5}$
were (seemimgly) guided by the results of lattice
calculations in which confinement/deconfinement $(c/d)$ and
chiral symmetry breaking/restoration $(b/r)$ transitions had been found to
coincide at $\mu =0.$
Moreover "deconfinement" was (and still is)
usually understood as the straightforward $H\leftrightarrow$
$QGP$ transition.
Accordingly, e.g. a condition was superimposed$^{\rm 5}$ on
pressures $p_{H}$ and $p_{q}$ of $H$ and $QGP$ phases at the transition
point (for $\mu =0): p_{H}(T_{c},0)=p_{q}(T_{c},0)$. Herefrom
the ratio $\beta=B_{q}/B_{Q}$
was obtained thus leaving only one of these
two parameters
undefined. In other works similar coincidence of $(c/d)$ and
$(b/r)$ at $\mu  =0$ was also taken as a basis. Herefrom, for a freely
 chosen $B_Q$, followed the value of $B_{q}$. All authors accordingly used
mainly the values $\beta  = B_{q}/B_{Q} \sim 3 \div 4$ with $B_{Q}$ more or
less close to the MIT bag value,
$B_{MIT} \simeq 56~{\rm MeV/fm}^{3} \quad
(~ B_{Q} = 56 \div 150~{\rm MeV}/{\rm fm}^{3}, \quad
B_{q} =200 \div 500~{\rm MeV/fm}^{3}$, keeping in all cases$^{\rm 2-5}$
$\beta  \sim 3 \div 4$ ~).
\par
Assumption concerning $B_{Q} \simeq  B_{MIT}$ seems to
be reasonable at least in the vicinity of the $H\leftrightarrow Q$ transition.
In fact, here valon density in $Q$ phase, at least in the beginning of
the valon
deconfinement process, is the same as in a nucleon (see Fig.$2b)$.
Accordingly we
shall use the assumption:
\begin{equation}
50 \leq  B_{Q} \simeq  B_{MIT} \leq  100~{\rm MeV/fm}^{\rm 3}.
\label{16}
\end{equation}
\par
On the contrary, argumentation which have lead to the
above mentioned estimate of $\beta =B_{q}/B_{Q}\sim 3$ is not admissible for
us since {\sl difference} between con\-fi\-ne\-ment/deconfinement $(c/d)$
of valons (which takes place with broken chiral
symmetry) and chiral symmetry breaking/restoration $(b/r)$ is
{\it exactly the phenomenon we are looking for}. Accordingly this
difference should not be excluded at the start of the
analysis as has been, as a matter of fact, done e.g. in$^{\rm 5}$.
Thus $B_{q}$ is a parameter which needs its
own physical foundation. Since it refers to $QGP$, it should
be tightly connected with the vacuum pressure $B_{\hbox{\sl vac}}~
\simeq  500 \div 1000\hbox{ MeV}/{\rm fm}^{3}$
and be close to it. Accordingly we are going to use larger
values of $\beta=B_{q}/B_{Q}$ putting it equal to $\beta \sim~10$.
Below we shall use the following set of reasonable
parameters as the {\sl standard} one:
\begin{equation}
B_{q}=500 \hbox{ MeV}/{\rm fm}^{3};\,\,\,B_{Q}=50
\hbox{ MeV}/{\rm fm}^{3};\,\,\,
r_{N}=0.8 {\rm fm};\,\,\, r_{Q}=0.3 {\rm fm}.
\label{17}
\end{equation}
\noindent The influence of $B_Q$ and $B_q$ as well as $r_j$ variations are
partially investigated.
{}~\\~\\~\\
{\bf MAIN RESULTS}
{}~\\
\par
Let us start with extreme cases: $\mu =0,T\neq 0 $(Figs.3,4) and
$T=0, \mu \neq 0 $(Fig.5).
\par
Fig.3 clearly demonstrates characteristic pattern of
double phase transition at the $p-T$ plane for chosen standard
parameter set (17).
It is evident that two
phase
transitions are well separated, $Q$ is the most stable one
(i.e. its pressure $p_{Q}$ is higher than that of other phases)
within the temperature interval of $\Delta T=T_{ch}-T_{d} \sim 50$ MeV,
with absolute values of transition temperatures being of the order
of the direct phase transition $(H\leftrightarrow QGP$) temperature $T_{c}$:
$T_{c} \simeq 180 MeV,\,\, T_{d} \simeq 145 Mev, \,\,T_{ch} \simeq 195 Mev.$
The value of this {\it "temperature corridor"} of $Q$ phase depends on $B_{q}$
(see Fig.4$(b)$) so that it becomes larger (up to 90 MeV)
for larger values of$B_q$ (when $B_{q}=1 \hbox{ GeV}/{\rm fm}^{3},
\beta \sim 20$,
 $T_{ch} \sim 230$ MeV) and shrinks to a
{\it triple point} (which are marked by the symbol "\#" within all the figures)
$T^{\#} \sim 140$ MeV for most frequently choosen
$B_{q} \simeq 200~{\rm MeV/fm}^3
\,\,\beta \sim 4$, close to that in$^{\rm 5}$. The same effect stays for
 $B_{Q}$ dependence: the closer it is to $B_{q}$, the
 narrower is $T$-region allowing for $Q$ phase existence, with the biggest
 value (corresponding to the triple point $T^{\#} \sim 180$ MeV)
 $B_{Q} \sim 160$ MeV (i.e. $\beta \sim 4.5$).
For all combinations of{\it B's} the chiral symmetry restores at $T_{ch}$
which is always larger than the critical temperature of direct transition
$T_{c}\equiv T_{H\leftrightarrow QGP}$ obtained in calculations when
ignoring existence of valons. Note that we do not show the influence of
r parameters becouse it is negligeable in this $T-\mu$ region.
\par
The similar pattern is presented in Fig.5 for $T=0, \mu \neq 0.$ for
the standard parameter set(17).
Here again there is a possibility for appearance of $Q$ phase region but now
it is greately suppressed: if $\beta =10$ while $B_{Q}$ has its minimal
reasonable value 50 MeV$/{\rm fm}^{3}$, the area covered by $Q$ phase on
$\{p,\mu \}$ diagramm is rather small and it shrinks into a triple point
already at
$B_{Q} \simeq 80 \hbox{ MeV}/{\rm fm}^{3}$; this region becomes
of course broader for larger $\beta$ values.
However it is essential that valons tell even at $T=0$. For more details
concerning $T=0$ see  $^{11}$.
\par
Now we go over to the general case of $T,\mu  \neq 0.$ The
$\{\mu ,T\}$ diagramm for the standard parameter set (17) is shown in Fig.6.
It is clear that there exists a wide corridor around the curve for direct
$H\leftrightarrow QGP$ transition which remains as well for various
combinations of parameters $(B_{q},B_{Q},r_{N}$, and $r_{Q})$ (see Figs.7).
It shrinks and even disappears with $\mu$ growth only at rather definite
conditions, namely for small $\beta $(the shape of $Q$ phase region
for $\beta <3 $ shown in Fig.7$(a)$, as well as in Fig.1,
reminds a thin banana) and for extremely large $r_{Q}$ (Fig.7$(b)$).
Otherwise transition from $H$ phase to $QGP$ proceeds only via valon phase
 (for detailes see  $^{\rm 12}$).
\par
Note that under not too large $\mu \leq 1$ GeV the master parameter
is $\beta
=B_{q}/B_{Q}$ while influence of particle interaction (i.e. choice of values of
$r_{N}$ and $r_{Q})$ is quite negligeable. Symbol line in Fig.7$(b)$
corresponds
to the case of all particles taken point like.
It differs considerably from the real particle diagram
only at $\mu \sim 1$ GeV, and covers the area having the shape of some
exotic fruit ("pineapple").
s\par
Critical values  $\epsilon^{cr}$ for energy densities (~that
is  the  ones on phase equilibrium curves~) are presented in Figs.8.
Let us point out (see Fig.$8(a)$) that the {\sl latent heat}, i.e. the jump
$\Delta \epsilon^{d} = (\epsilon^d_Q-\epsilon^d_H)$  at the
deconfinement transition is much smaller than that at
the chiral one, $\Delta \epsilon_{ch} = (\epsilon^{ch}_{q}-\epsilon^{ch}_{Q})$,
and much smaller than that for direct transition ignoring valons (Fig.$8(b)$).
The decomfinement transition requires rather "soft" experimental conditions.
{}~\\~\\~\\
{\bf DISCUSSION AND SUMMARY}
{}~\\
\par
Of course, obtained numerical values, in particular those of $T_{d}$
and $T_{ch}$, should not be taken literary.
E.g., they essentially depend on $B_Q$ and
$B_{q}$ values chosen in a rather rough manner. According to e.g.
Shuryak $^{\rm 13}$, for hadronic substance at $T=0,\,\,B_{eff}$
(efficient $B_{eff}$ describing the summarized influence of interaction
of particles between themselves and the vacuum)
is not reducible to two constants but changes continuously increasing
with baryon number density $n_{B}$ from some lower value of the order
of $B_{MIT}$ up to $B_{eff} \sim B_{vac}$ for $QGP$.
However, our calculations $^{\rm 12}$ show that baryon number density in
$Q$ phase, $n_{Q}$, is much closer to that in the $H$  phase  than
to the one in the $QGP$. Thus the use of only two values - $B_{Q}$
and $B_{q}$ -
seems to be not too bad.
\par
Further, our calculations show that particle density of $H$
phase in the vicinity of transition curve to $Q$ phase at relatively
large $\mu >1$ GeV is very close to its limiting value $\sim v^{-1}_{N}$
(close packing) and use of the Van der Waals gas approximation for it is very
suspicious. But the same weak point is met in the papers which lead
to pessimistic conclusions concerning possibility of the $Q$-phase$^{\rm 2-4}$.
Repulsive potential for nucleons considered in $^{\rm 5}$ seems of
course to be
more realistic; similar potential will be discussed in our paper $^{\rm 12}$.
This does not change the final result qualitatively.
\par
Nevertheless qualitative and even semiquantitative
results seem to deserve attention. Comparing Figs.6,7 with
Fig.1 we see that for sufficiently large values of $B_{q}/B_{Q}$ the
pattern declared in the beginning of this paper (Fig.2) is almost fully
supported. We may conclude that there actually exists possibility
of hadronic phase to $QGP$ transfer (at seemingly reasonable choice
of parameters) to proceed only via an intermediate valon, i.e.
constituent quark plus pions, phase. Therefore, we expect existence of
two phase transitions. For $\mu =0$ transition temperatures differ by some
50 MeV.
One of them, deconfinement of valons, proceeds at the
{\it Hagedorn temperature} $T_{d}=T_{H}$, above which, in the deconfined valon
phase, hadrons exist but as a small thermal admixture. The second phase
transition at $T_{ch} \sim 200$ MeV corresponds to deconfiment/confinement of
current quarks and gluons and, simultaneously, restoration/breaking of
chiral symmetry. Thus difference between $T_{d}$ and $T_{ch}$ is
physically meaningful.
\par
The $Q$ phase with deconfined constituent quarks (valons)
having mass $\sim$ 320 MeV, due to $m_{Q}/T >1,$ is nonrelativistic,
its state equation is that of nonrelativistic Van der Waals gas
(since $r_{Q}$ is small). It is only in the vicinity of the phase transition
to $QGP$ that valon size begins to tell. However, this last transition at not
too large $\beta = B_{q}/B_{Q}$ takes place long before close packing of
valons is
attained$^{\rm 12}$. This result is rather interesting and has more
physical sense than had been put into the model. In fact the
model of hard black balls, being quite sufficient for nucleons due to
their high stability, is hardly valid for constituent quarks which may
become partialy transparent and larger. Overlapping of
valon tails securing exchange of current quarks and gluons between valons
may happen at distances exceeding the assumed black ball radius. Thus
the transition to $QGP$ phase does not necessarily require for
close packing of valons. This is not so for nucleons at
$H\leftrightarrow Q$ transition where the transition happens long after the
considerable close packing is attained; this may be ascribed
to necessity of overcoming the surface tension of nucleons.
However for more detailed analysis it is desirable to consider more
realistic repulsive potential for interaction of valons
than simple hard core model (e.g. at the same manner as it has been
done in$^{\rm 14}$).
\par
The obvious first application of these results of
course is to be to the evolution of matter formed at collision of highly
relativistic heavy nuclei. Let us stress two most
important items.
\par
First, as it have been shown above (Figs.8) the latent
heat value for $H\leftrightarrow Q$ transition is much smaller than that for
the "upper" transition $Q\leftrightarrow QGP$ (and much smaller than
people had obtained for direct, i.e. $H\leftrightarrow QGP$ transition). The
same is true$^{\rm 12}$ for transition pressure and $n_{B}$ values.
This means that first deconfinment transition may proceed under much
more "soft"
conditions than it was expected for direct transition $H\leftrightarrow QGP$:
it is sufficient to compress the nucleus only 3-4 times applying not very
high pressure. Thus this transition may happen at not too high energies of
colliding nuclei, i.e. {\sl even at Bevalac or in Dubna$^{\rm 7,8}$} where
appearance of
pure $QGP$ is hardly possible. Accordingly, e.g.in the Dubna experiments
one may
expect to observe some effects impossible for simple $pp$ collisions.
In particular, our estimates show that admixture of $\Lambda $ particles
may attain  in central rapidity region some 10 percents of nucleons and
this reminds us Okonov group Dubna experiment $^{\rm 15}$ in which
$\Lambda 's$ were especially studied.
\par
Secondly, transition pattern described above should influence essentially
the character of hydrodynamical expansion and cooling of initially hot
matter both if it takes place either in $Q$ or in $QGP$ phase. For
$QGP$ initial state the following picture seems to be natural. The system
experiences the phase transition to $QGP$ phase. After cooling down to
chiral transition temperature $T_{ch} \sim 200 MeV$, it would further expand
and cool down within a large temperature interval according to
nonrelativistic state equation.
Since at present (~and later when RHIC will be operating~) the attainable
initial temperature may exceed $T_{ch}$ but slightly, the time
spent for getting $T_{ch}$, as well as duration of cooling down in $H$
phase from $T_{d} \sim 150 MeV$ to freezing temperature,
$T_{f} \sim 140$ MeV, may be smaller than the time of existence of
$Q$ phase with its nonrelativistic state equation. Therefore the
existence of $Q$ phase should tell itself essentially. Here
detailed calculations for resulting rapidity distribution, etc.
are necessary to compare with experiment (at present they are in progress).
\par
The results obtained seem to contradict to the statement based at
lattice calculations that at $\mu =0$ two phase transitions
are to coincide. The situation here is however disputable.
Although existence of $Q$ phase was seemmgly never especialy looked for in
lattice calculations there are some  works in which existence of "heavy modes"
was assumed in order to get rid of some results looking  physically
unsatisfactory$^{\rm 16}$ (~the  $QGP$ phase does not have properties of an
ideal relativistic gas~). Lattice specialists should answer the question
whether the double phase transition with $Q$ mass at 320 MeV could be
noticed in lattice calculations performed until now and how it
should show itself.
{}~\\~\\~\\
{\large {\bf Acknowledgments}}
{}~\\
\par
The authors wish to express their gratitude for critical remarks by
I.V.Andreev, I.M.Dremin and  I.I.Royzen.  We  are  also indebted  to
G.M.Zinoviev and M.Polikarpov for fruitfull discussions.
The work was supported partly by
the Russian Foundation  for Fundamental Researches, grants Nos. 94-02-15558 and
94-02-3815.\\
{}~\\~\\
{\bf FIGURE CAPTION.}\\
{}~\\
{\bf Figure~1.} Phase transition curves in the $T-\mu$ plane(from$^5$):
$Q$-phase region is between dashed and dotted lines.\\
{}~\\
{\bf Figure~2.} Schematic picture of nuclear matter transformation
under compression.\\
{}~\\
{\bf Figure~3}. Phase diagram for $\mu=0$ case at the ${P-T}$ plane for
standard parameter set(17): the solid, dashed and dotted lines correspond
to the the pressures of hadronic $(H)$, valonic $(Q)$ and current $q(QGP)$
phases respectively.\\
{}~\\
{\bf Figure~4.} The same plot as in Fig.3 for standard parameter
set (solid lines) and varying {\it B's}: {\bf (a)} - $B_Q \simeq 100$ (dashed
line), 160 MeV/fm$^3$(dotted line); {\bf (b)} -$B_q \simeq 1000$ (dashed
line), $200$ MeV/fm$^3$ (dotted line);"\#" marks the triple points.\\
{}~\\
{\bf Figure 5.} Phase diagram in the $p-\mu$ plane for T=0 and
standard parameterset; designations are the same as in Fig.3.\\
{}~\\
{\bf Figure~6.} Phase transition curves in the $T-\mu$ plane: valon
deconfinement (dashed line), direct transition ignoring valons (solid
line), chiral transition (short dashed).\\
{}~\\
{\bf Figure~7.} Transition curves bounding the $Q$-phase region: {\bf (a)} -
for standard $r_j$ and $B_q$, $B_Q= 50$ MeV/fm$^3$ (solid lines),
100 MeV/fm$^3$ (long-dashed lines) and 150 MeV/fm$^3$ (dotted lines);
short-dashed line
corresponds to $B_q=1$ GeV/fm$^3$; {\bf (b)} - for standard $B$'s and varying
$r_Q= 0.2$ fm (solid lines), 0.4 fm (dashed lines) and $r_j=0$
($\dagger$ lines).\\
{}~\\
{\bf Figure~8.} Energy densities $\epsilon_j^{cr}$ curves for: {\bf (a)}
 $H$ and $Q$ matter at deconfinement transition (solid lines), $Q$ and $QGP$
 matter at chiral transition (dashed lines); {\bf (b)} $H$ and $QGP$ matter
at direct transition ignoring valons
(dotted lines).\\~\\

{\bf REFERENCES}\\
{}~\\
\begin{tabular}{p{4mm}p{150mm}}
1.&B.L.Ioffe,V.A.Khoze, L.N.Lipatov, "Hard Processes". Volume 1.
"Phenomenology, Quark-Parton Model", North Holland, Amsterdam, 1984
\end{tabular}
\\
\begin{tabular}{p{4mm}p{150mm}}
2.&J.Cleymans, K.Redlich,H.Satz,E.Suhonen Z.Phys.C 33(1986), 151
\end{tabular}
\\
\begin{tabular}{p{4mm}p{150mm}}
3.&S.Sohlo, E.Suchonen J.Phys.G.Nucl.Phys. 13(1987), 1487
\end{tabular}
\\
\begin{tabular}{p{4mm}p{150mm}}
4.&Kuono H., Takagi F. Z. Phys. C 42 (1989), 209
\end{tabular}
\\
\begin{tabular}{p{4mm}p{150mm}}
5.&D.V.Anchishkin, K.A.Bugaev, M.I.Gorenshtein, E.Suhonen
Z.Phys.C45(1990)687
\end{tabular}
\\
\begin{tabular}{p{4mm}p{150mm}}
6.&E.V.Shuryak Phys.Lett. 107B(1981), 103;
Nucl.Phys. B203 (1982), 140
\end{tabular}
\\
\begin{tabular}{p{4mm}p{150mm}}
7.&E.Feinberg On Deconfinement of Constituent and Current
Quarks in  Nucleus-Nucleus  Collisions  P.N.Lebedev
Institute  Preprint, 1989, N 177.
\end{tabular}
\\
\begin{tabular}{p{4mm}p{150mm}}
8.&E.Feinberg  in  Relativistic  Heavy  Ion  Collision,  ed.  by
L.P.Czernai and D.D.Strottman, World Scientific, 1991 (see Chapter
5, Section 7)
\end{tabular}
\\
\begin{tabular}{p{4mm}p{150mm}}
9.&R.Hagedorn, J.Rafelsky, in:Thermodynamics of quarks and hadrons.
H.Satz(Ed.). Amsterdam: North Holland 1981; R.Hagedorn:Z.Phys.C - Particles and
Fields 17, (1983), 265
\end{tabular}
\\
\begin{tabular}{p{4mm}p{150mm}}
10.&V.V.Anisovich,M.N.Kobrinsky,   Y.Niri
and   Y.M.Shabelski, Sov.Phys. Uspekhi, 4  (1984) 553
(see section 1)
\end{tabular}
\\
\begin{tabular}{p{4mm}p{150mm}}
11.&O.D.Chernavskaya,
"Double   phase   transition   at   zero temperature" (to be published)
\end{tabular}
\\
\begin{tabular}{p{4mm}p{150mm}}
12.&O.D.Chernavskaya, E.L.Feinberg,
"The possibility of double phase transition via the deconfined
constituent quark phase" (to be published)
\end{tabular}
\\
\begin{tabular}{p{4mm}p{150mm}}
13.&E.V.Shuryak The QCD Vacuum, Hadrons and the Superdense
Matter, World Scientific, 1988, see fig.(9.2)
\end{tabular}
\\
\begin{tabular}{p{4mm}p{150mm}}
14.&Ch. Barter,D.Blaschke,H.Voss, Phys.Lett. B 293 (1992) 423
\end{tabular}
\\
\begin{tabular}{p{4mm}p{150mm}}
15.&Anikina M.et.al. Z.Phys. C25 (1984)1;Phys.Rev. C 23 (1986) 895.
Okonov E.O. in: Proceed. of Intern. Symposium on Modern Developments in Nuclear
Physics, Novosibirsk, 1987, p.166, World Scientific
\end{tabular}
\\
\begin{tabular}{p{4mm}p{150mm}}
16.&F.Karsch Prep. CERN TH/13/94
\end{tabular}
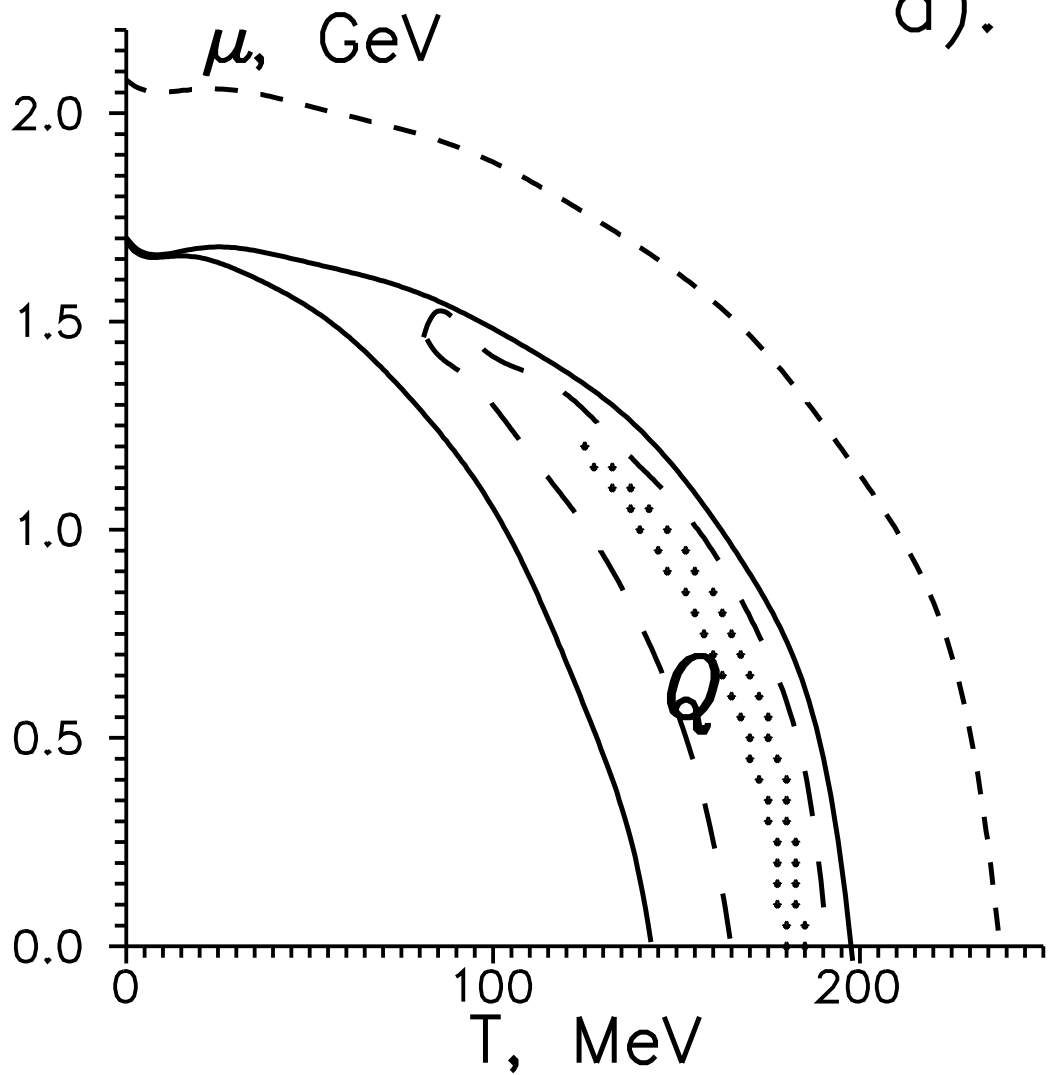
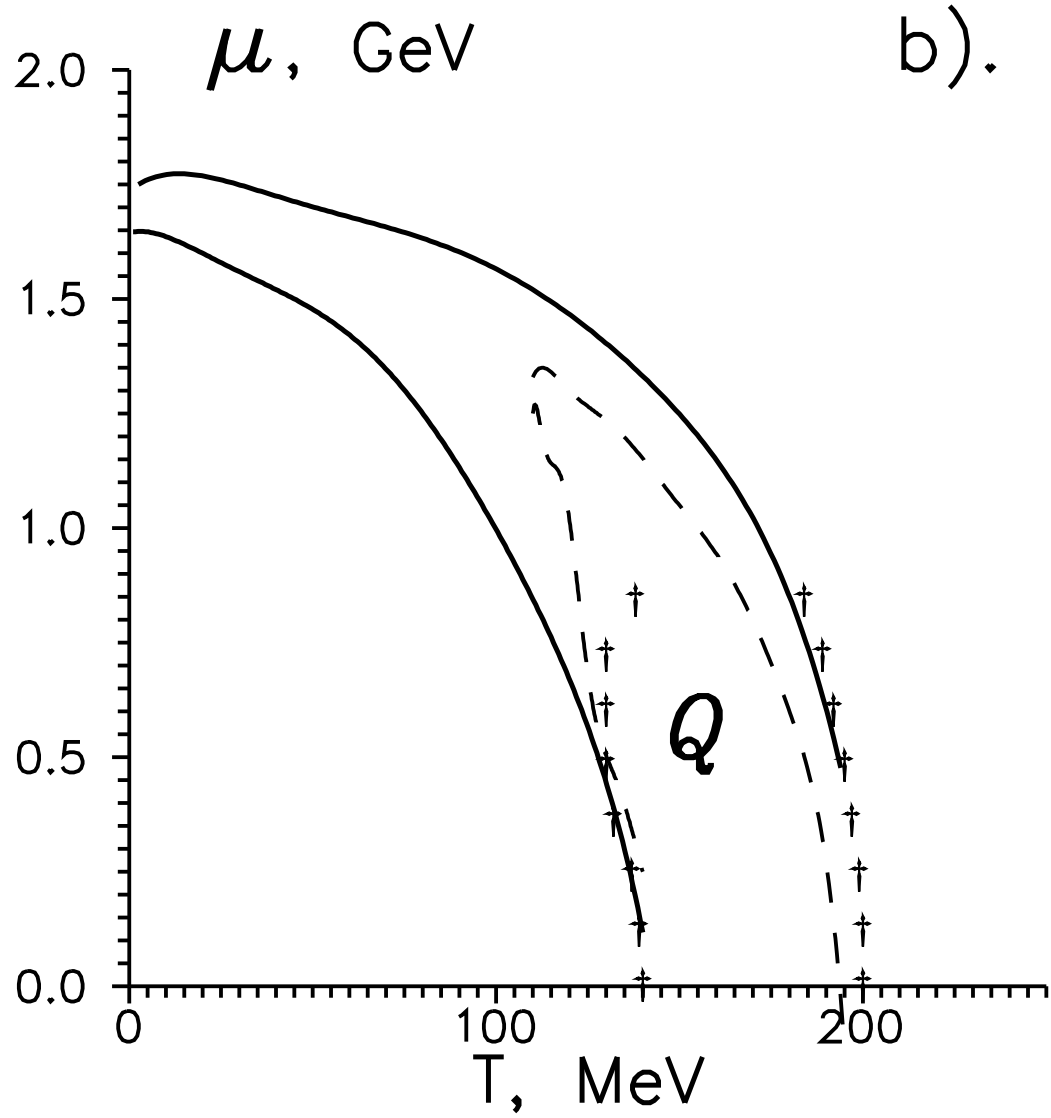
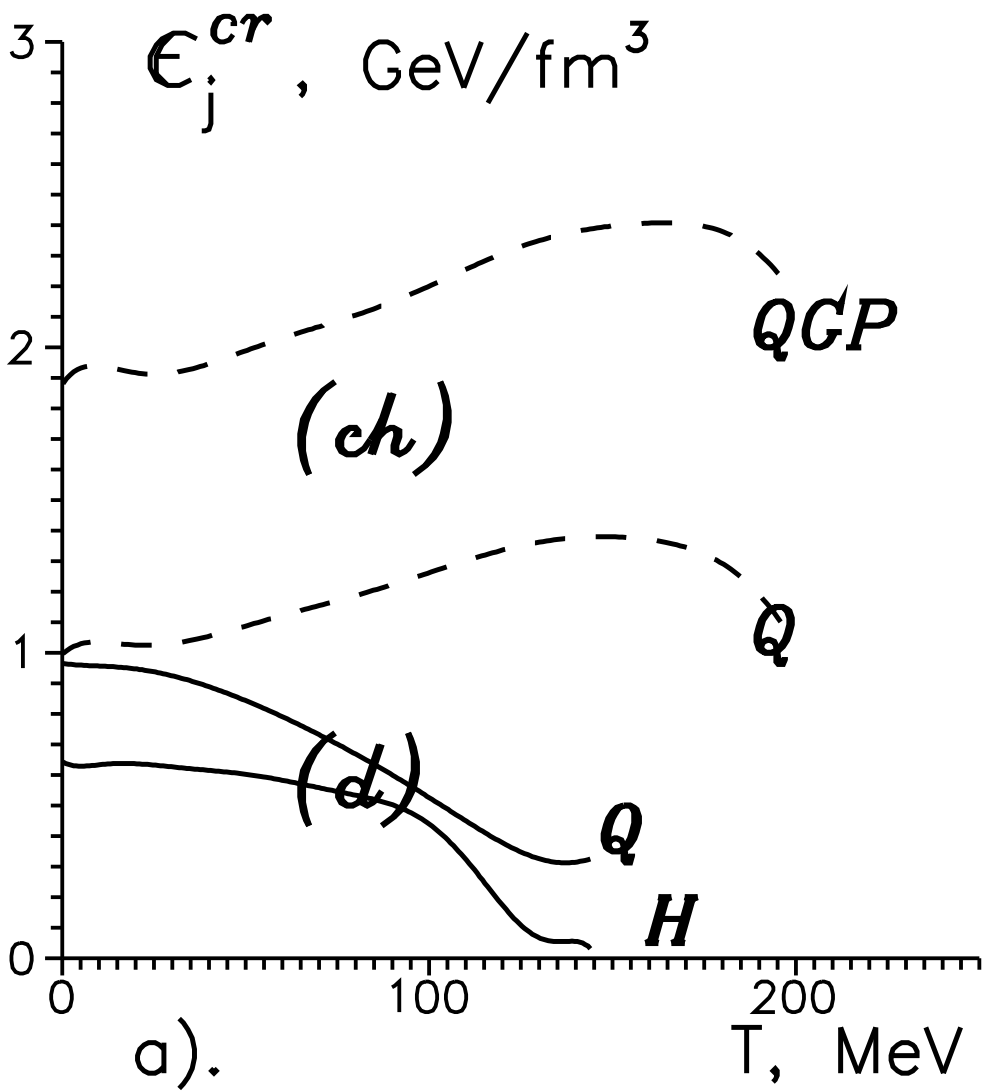
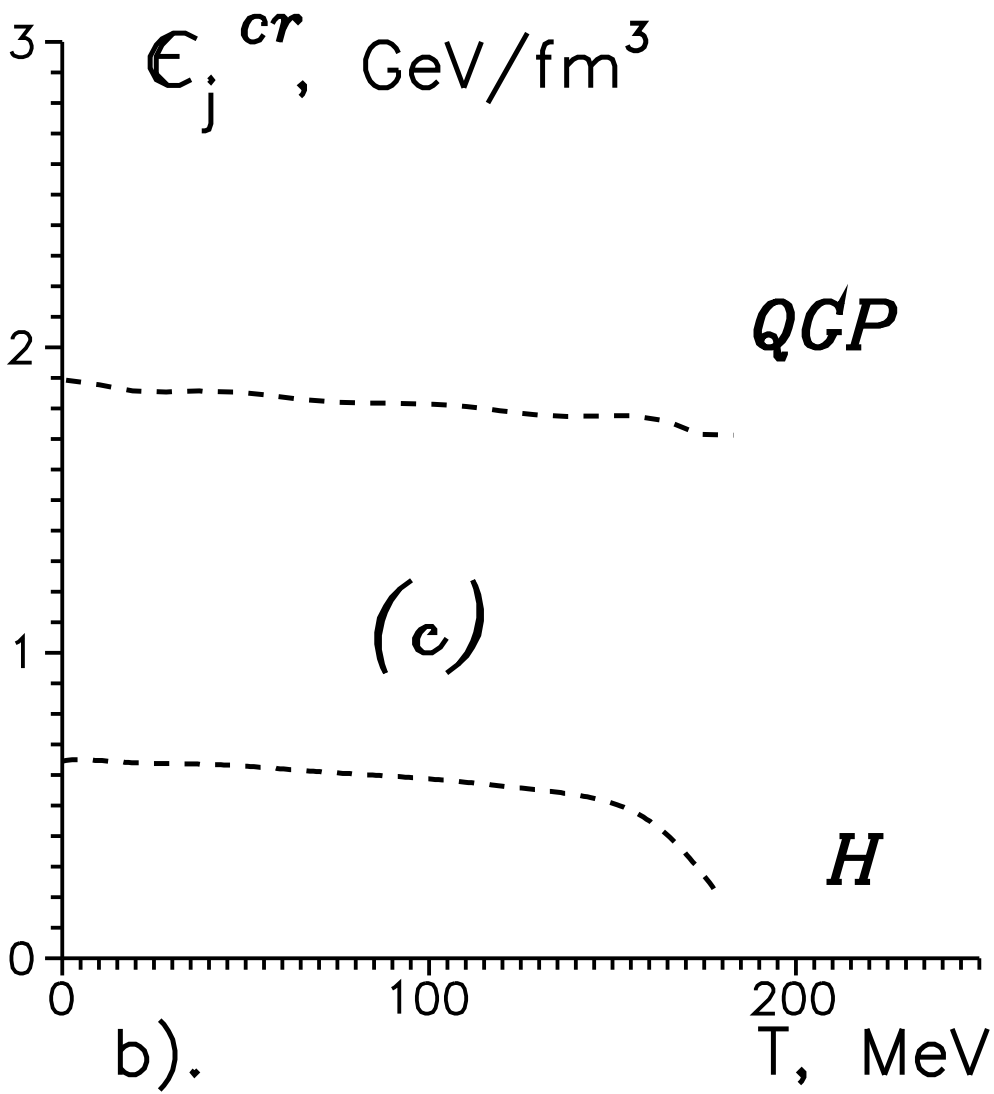
\end{document}